\documentclass[prd,twocolumn,nofootinbib,amsmath,amsfonts,amssymb,showpacs,preprintnumbers]{revtex4-1}

\usepackage[usenames]{color}
\usepackage{epsfig,graphicx,bm,color}

\usepackage{natbib}
\newcommand{\citefig}[1]{Fig.~\ref{#1}}

\newcommand{\eg}{{\it e.g.}}

\newcommand{\ben}{\begin{eqnarray}}
\newcommand{\een}{\end{eqnarray}}
\newcommand{\be}{\begin{equation}}
\newcommand{\ee}{\end{equation}}

\newcommand{\mchi}{\mbox{$m_{\chi}$}}
\newcommand{\sigv}{\mbox{$\langle \sigma v \rangle $}}

\newcommand{\aj}{Astron. J.}
\newcommand{\aap}{Astron. Astroph.}

\newcommand{\jgr}{J. Geophys. Res.}
\newcommand{\jcap}{JCAP}
\newcommand{\mnras}{MNRAS}
\newcommand{\physrep}{Phys. Rept.}

\graphicspath{{./Figures/}}

\begin{document}

\title{10 GeV dark matter candidates and cosmic-ray antiprotons}

\author{Julien Lavalle}
\affiliation{Dipartimento di Fisica Teorica, Universit\`a di Torino \& INFN,
  via Giuria 1, 10125 Torino --- Italy}
\email{lavalle@to.infn.it}
\email{current address: IFT-Madrid, Spain}

\date{12 July 2010}

\begin{abstract}
Recent measurements performed with some direct dark matter detection 
experiments, \eg~CDMS-II and CoGENT (after DAMA/LIBRA), have unveiled a few 
events compatible with weakly interacting massive particles. The preferred mass 
range is around 10 GeV, with a quite large spin-independent cross section of 
$10^{-43}$-$10^{-41}\,{\rm cm^2}$. In this paper, we recall that a light dark 
matter particle with dominant couplings to quarks should also generate 
cosmic-ray antiprotons. Taking advantage of recent works constraining 
the Galactic dark matter mass profile on the one hand and on cosmic-ray 
propagation on the other hand, we point out that considering a thermal 
annihilation cross section for such low mass candidates very likely results in 
an antiproton flux in tension with the current data, which should be taken
into account in subsequent studies.
\end{abstract}

\pacs{95.35.+d, 95.30.Cq, 96.50.S, 12.60.-i}

\maketitle

\preprint{DFTT 10/2010}
The DAMA Collaboration has long claimed the detection of an annual 
modulation in their data~\cite{2003NCimR..26a...1B,2008EPJC..tmp..167B}, which, 
if interpreted in terms of dark matter interaction with the detector, seems to 
favor weakly interacting massive particles (WIMPs) with light masses 
(see \eg~\cite{1999PhRvD..59i5004B}). More recently, the 
CDMS-II~\cite{2009arXiv0912.3592T} and
CoGENT~\cite{2010arXiv1002.4703A} Collaborations, have announced excess events 
in their data. Interpretations in terms of dark matter were performed in 
\eg~\cite{2010PhRvD..81j7302B} and \cite{2010JCAP...02..014K} (see also
\cite{2010arXiv1006.3318M} for less conventional models), indicating that
WIMPs with masses around 10 GeV could also explain these measurements. In 
Ref.~\cite{2010JCAP...02..014K}, it was notably shown that some of the favored 
regions may not be compatible among each other.

Interesting constraints on such light WIMPs may actually come from colliders
\cite{2010arXiv1005.1286G,2010arXiv1008.1783G,2010arXiv1005.3797B}.
From the astrophysical point of view, the annihilation of such light WIMPs may 
generate high gamma-ray fluxes which are at the edge of exclusion with current 
measurements~\cite{2010arXiv1007.2765A}, but there are still large uncertainties
coming from our incomplete knowledge of the detailed dark matter distribution, 
in particular in the centers of galaxies. On the Galactic scale, cosmic-ray 
antiprotons also provide interesting constraints~\cite{1984PhRvL..53..624S} 
since predictions are less sensitive to the choice of the halo profile, but more
to the local density (see \eg~\cite{2004PhRvD..69f3501D}) --- indeed, the 
relevant annihilation yield is averaged over a volume set by the diffusion 
scale, which of the order of a few kpc about the Earth. It was already noticed 
in \cite{2005PhRvD..72h3518B} that light neutralinos with masses below 10 GeV 
might generate antiproton fluxes overshooting the data if the dominant 
annihilation proceeds into $b\bar{b}$, because the $1/\mchi^2$ flux suppression
is no longer efficient with respect to heavier WIMPs. In their analysis, these 
authors used a smooth cored isothermal halo profile for the dark matter 
distribution in the Galaxy, which was not meant to be in clean agreement with 
the Galactic rotation curves. Nevertheless, the largest uncertainties did 
actually come from propagation, notably from the relative freedom in setting the
vertical extent $L$ of the diffusion zone. Indeed, in their minimal case of 
$L=1$ kpc, at the pessimistic edge of cosmic-ray nuclei 
constraints~\cite{2001ApJ...555..585M,1998ApJ...509..212S}, the dark matter 
contribution was shown to be dramatically decreased by almost 1 order of 
magnitude with respect to the best-fit propagation setup.

\begin{figure*}[t!]
 \centering
\includegraphics[width=\columnwidth]{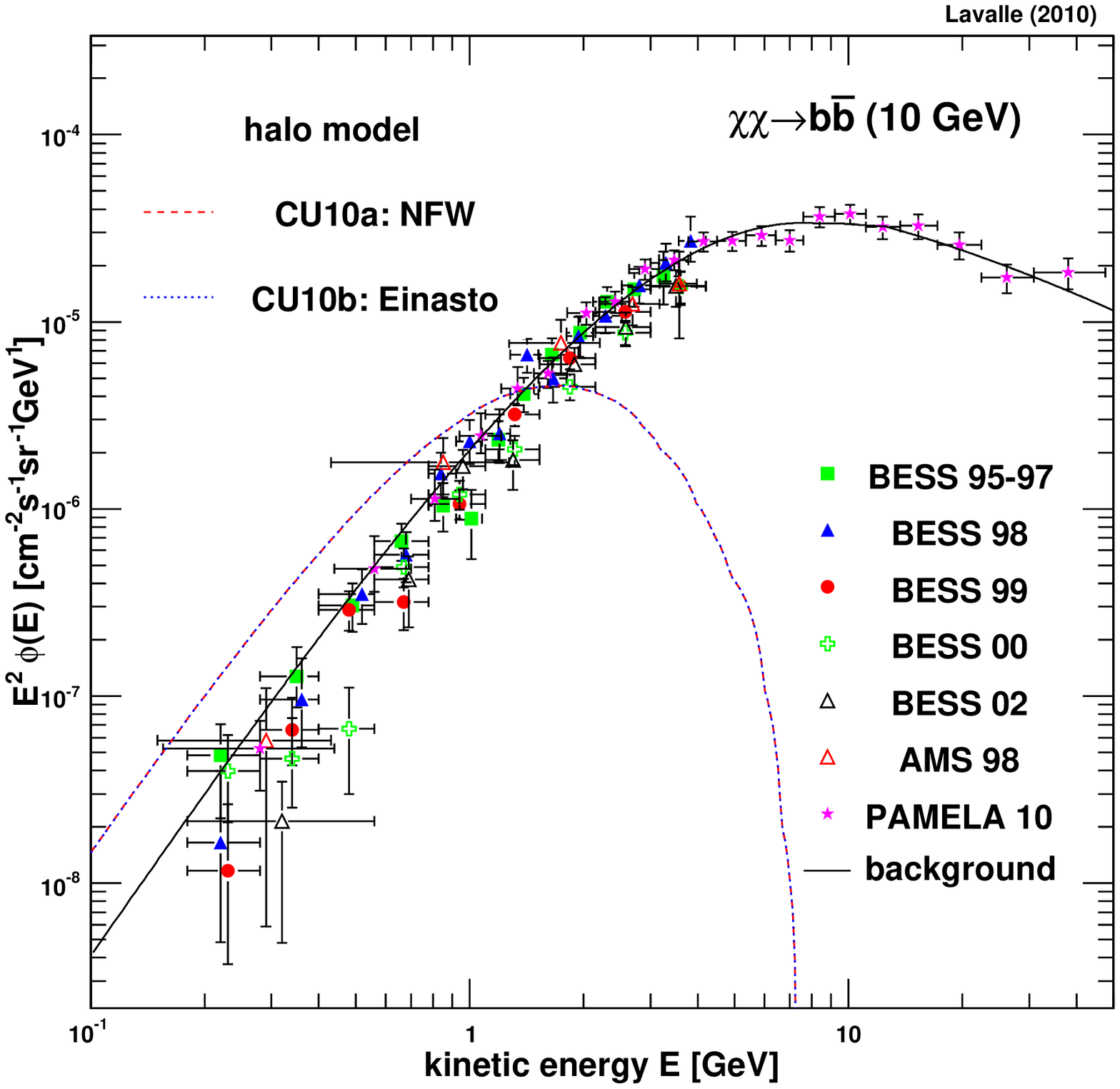}
\includegraphics[width=\columnwidth]{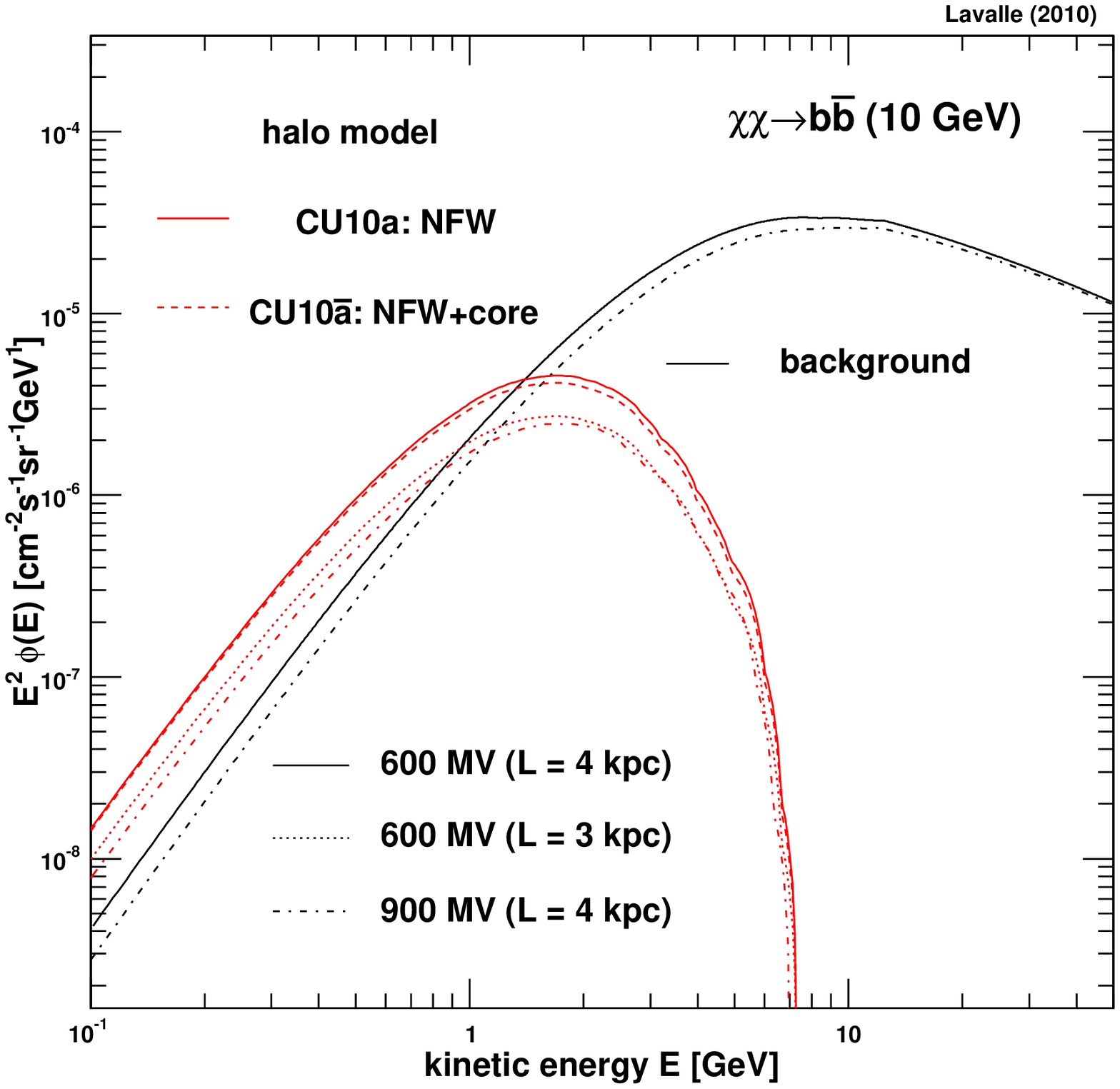}
\caption{Left: Predictions of the primary antiproton flux for a 10 GeV WIMP
  annihilating into $b\bar{b}$ pairs, for the NFW (CU10a) and the Einasto 
  (CU10b) profiles derived in~\cite{2010JCAP...08..004C}. The solid black curve 
  is the secondary background predicted using the same propagation setup
  \cite{2001ApJ...563..172D}. Right: Impact (i) of imposing a core to the NFW 
  case (CU10a versus CU10$\bar{\rm a}$, dashed curve), (ii) of increasing the 
  solar modulation force field (dotted-dashed curves), and (iii) of decreasing 
  the vertical halo boundary $L$ from 4 to 3 kpc (dotted curve).}
\label{fig:results}
\end{figure*}

Here we take advantage of the recent works performed (i) by Catena and Ullio
\cite{2010JCAP...08..004C} (CU10 hereafter) on constraining the Galactic dark 
matter distribution from kinematic data on the one hand, and (ii) by Putze, 
Derome, and Maurin~\cite{2010A&A...516A..66P} (PDM10 hereafter) on cosmic-ray 
propagation on the other hand, to improve the antiproton analysis.

CU10 notably showed that one could reach interesting constraints on the
local dark matter density by using updated tracers of the Galactic dynamics,
provided some initial assumptions about the dark matter profile. These 
assumptions can be made on well-motivated theoretical grounds, since the 
highest-resolution cosmological N-body simulations to date focused on 
Milky-Way-like galaxies now seem to converge towards similar predictions, in 
between an Einasto~\cite{1968PTarO..36..357E,2006AJ....132.2685M} and an 
Navarro-Frenk-White \cite{1997ApJ...490..493N} (NFW) profile 
(\eg~\cite{2008Natur.454..735D,2008Natur.456...73S,2010MNRAS.402...21N}).
Using these assumptions, CU10 derived a local dark matter density
of $\rho(r_\odot=8.25\pm 0.29 \,{\rm kpc}) = 0.386\pm 0.027\,{\rm GeV/cm^3}$ 
in the former case and of $\rho(r_\odot=8.28\pm 0.29 \,{\rm kpc}) = 
0.389\pm 0.025\,{\rm GeV/cm^3}$ in the latter case. We note that these
results for the local density were confirmed independently 
by~\cite{2010arXiv1003.3101S}, with a completely different method --- these
authors derived $\rho_\odot = 0.41\pm 0.11\,{\rm GeV/cm^3}$. There are still,
obviously, large uncertainties with respect to the dark matter distribution
in the inner kpc about the Galactic center, since either the baryons and
the central black-hole may play important roles there, increasing or decreasing 
the inner density depending on the 
hypotheses~\cite{1999PhRvL..83.1719G,2010Natur.463..203G}. Nevertheless,
we underline that decreasing the density in the inner kpc has much less impact 
on the antiproton flux predictions than on the gamma-ray flux predictions 
because of spatial diffusion, as it is demonstrated further below. This is 
strengthened by the fact that the (anti)proton propagation scale 
increases with energy: antiprotons are more local if belonging to the energy 
range of interest here, say below 10 GeV, than those of higher energies. 
Typically, the propagation scale $\lambda$ is set by the ratio of diffusion to 
convection and spallation, and is of order of a very few kpc for a 1 GeV 
antiproton, whatever $\lambda<L$~\cite{2003A&A...402..971T}. Otherwise, $L$
provides an extra-limit to the propagation scale: the probability to escape
the diffusion halo strongly increases for $\lambda\gtrsim L$, so that 
antiprotons can hardly come from regions distant by more than a very few
times $L$.

Concerning uncertainties in the cosmic-ray propagation parameters, $L$, 
as we stressed above, has the most dramatic impact on dark matter signals.
Nevertheless, the recent MCMC analysis performed by PDM10 showed that
adopting a very small diffusive halo with $L$ down to 1 kpc makes it very 
difficult to accommodate the current cosmic-ray nuclei data. Moreover, large 
diffusion halo models, as large as $\sim 10$ kpc, are also preferred in 
astrophysical studies of the high-latitude diffuse gamma-ray emission measured 
by Fermi~\cite{2009ApJ...703.1249A} (see~\eg~\cite{porter_ppc10}). This
minimal $L$ of 1 kpc proposed in \cite{2004PhRvD..69f3501D} was actually
not really motivated by observational facts, but by the sake of a very 
conservative approach. As independent hints in this respect, it is 
indeed commonly observed that the radio halos of nearby spiral galaxies have 
sizes larger than 1 kpc (\eg~\cite{1995A&A...302..691D}), and a few kpc 
vertical extent is also favored by studies of the Galactic magnetic field 
(\eg~\cite{2008A&A...477..573S}).

In the following, we adopt the best-fit model derived 
in~\cite{2001ApJ...555..585M}, which is astonishingly close to the best-fit
setup derived in PDM10, in which the diffusion halo has a vertical extent of 
$L=4$ kpc, still much lower than what is suggested by the diffuse gamma-ray 
interpretation. We study a 10 GeV dark matter particle candidate, Majorana 
fermion or scalar, with a thermal annihilation cross section 
$\sigv=3\times 10^{-26}{\rm cm^3/s}$, entirely annihilating into $b\bar{b}$ 
quark pairs --- using other quark flavors would barely change the antiproton 
production; furthermore, taking a 
lower branching ratio into quarks would translate linearly into our analysis 
results. We have derived the injection antiproton spectrum with the public code 
PYTHIA \cite{2006JHEP...05..026S}. For the dark matter profile, we consider the 
NFW and Einasto profiles constrained in CU10 and discussed above, which we 
respectively denote CU10a and CU10b in the following. We also investigate the 
potential effect of imposing a 1 kpc core to the NFW case, which we denote 
CU10$\bar{\rm a}$.

In \citefig{fig:results}, we show the antiproton flux predictions at the Earth
obtained with the series of ingredients introduced above. We also plot the 
secondary antiproton background consistently derived within the {\em same} 
propagation setup~\cite{2001ApJ...563..172D}. The data points are taken from
\cite{2000PhRvL..84.1078O,2001APh....16..121M,2002PhRvL..88e1101A,2005ICRC....3...13H,2002PhR...366..331A,2010arXiv1007.0821P}. In the left panel, it clearly 
appears that the difference coming from using different halo profiles is 
negligible --- at the order of a few percents, hard to see from the plot. This 
comes from the fact that the local normalization of the dark matter density is 
quite the same, and that the global shape does not differ significantly on the 
kpc scale around the Earth. We note that the primary contribution originating 
from dark matter annihilation does exceed the secondary background\footnotemark 
~below $\sim 2$ GeV, overshooting it by a factor up to almost 5 around 200 MeV. 
In the right panel, we illustrate the effect of imposing a core to the NFW case 
(CU10$\bar{\rm a}$, dashed curve), and demonstrate that this has a very 
poor impact. This is due to the limited propagation scale that characterizes the
transport of low-energy antiprotons. We also study the influence of modifying 
the force field applied for solar modulation~\cite{1971JGR....76..221F}, 
increasing it from $\phi = 600$ MV (solid curves) to 900 MV (dotted curves). 
Such a change applies to both the signal and the secondary background, which 
makes the argument still valid in the strong solar activity regime. 
Finally, to allow a more conservative view, we consider a decrease of $L$ 
down to $3$ kpc self-consistently with the cosmic-ray nuclei constraints
\cite{2001ApJ...563..172D} (dotted curve) --- such a change has no effect
at all on the background prediction. We see that even in that
case, the predicted primary flux exceeds the secondary background by a factor
of 2, leading again to serious tensions with the data. We still emphasize that
many independent hints favor a large diffusion halo model with $L>3$ kpc, 
as already mentioned above.

\footnotetext{For the non-expert reader, {\em secondary} means  that the 
  antiproton background comes from {\em secondary} astrophysical processes, 
  namely nuclear interactions of standard cosmic-ray nuclei (mostly protons) 
  with the interstellar gas (mostly hydrogen).}

Therefore, a light dark matter particle in the $\sim$ 10 GeV mass range, 
annihilating into quark pairs at the thermal rate set by the relic density 
constraints, is expected to harden the antiproton spectrum below a few GeV. 
This very likely leads to important tensions with the current data. Further 
accounting for the predicted presence of subhalos would make this statement
even slightly more severe~\cite{2009arXiv0908.0195P}, as well as considering
a dark matter disk due to subhalo tidal streams trapped into the Galactic
disk~\cite{2009ApJ...703.2275P,2009MNRAS.397...44R,2010PhRvD..82b3534L}. 
Turning the argument around, observing a net inflexion in the antiproton 
spectrum around a few GeV could be a hint pointing towards the contribution of 
light WIMPs --- it does not seem to be the case in the available data. A 
loophole is of course possible so as to decrease the primary signal and escape
these constraints: \eg~combining either a lower branching ratio into quarks or 
a lower annihilation cross section with playing with the astrophysical 
parameters, or, for complete safety, taking a WIMP mass less than the 
(anti)proton mass~\cite{2010arXiv1008.1784C}. Concerning the astrophysical 
parameters, the ongoing efforts to constrain them have started to allow for 
less freedom in the predictions, at least in the domain of local charged cosmic 
rays.

To conclude, we emphasize that light dark matter candidates considered
in the interpretation of direct detection signals should be checked against
the cosmic-ray antiproton data, at least whenever their couplings to quarks
are significant. A more systematic study of this complementarity is on-going for
different particle physics scenarios beyond the standard model and is about
to show that some models are already excluded, except in contrived 
astrophysical situations~\cite{cerdeno_inprep}.

{\bf Acknowledgements:} It is a pleasure to thank D. Maurin for interesting
discussions about the diffusion halo size. We are also grateful to D. 
Cerde\~no and E. Nezri for stimulating exchanges about direct dark 
matter searches, and for complementary on-going works. We finally kindly thank 
R. Catena for providing us with details on his dark matter mass profile 
analysis.

\bibliography{lavalle_bib}


\end{document}